\documentstyle[aps,multicol]{revtex}

\title{Algorithm for numerical integration of the rigid-body
       equations of motion}

\author{Igor P. Omelyan \\[3pt]}

\address{Institute for Condensed Matter Physics,
         National Ukrainian Academy of Sciences, \\
         1 Svientsitsky st., UA-290011 Lviv, Ukraine.
         E-mail: nep@icmp.lviv.ua \\[4pt]}

\begin{document}

\maketitle

\begin{abstract}

A new algorithm for numerical integration of the rigid-body equations
of motion is proposed. The algorithm uses the leapfrog scheme and the
quantities involved are angular velocities and orientational variables
which can be expressed in terms of either principal axes or quaternions.
Due to specific features of the algorithm, orthonormality and unit norms
of the orientational variables are integrals of motion, despite an
approximate character of the produced trajectories. It is shown that
the method presented appears to be the most efficient among all known
algorithms of such a kind.

\vspace{10pt}


\end{abstract}

\vspace{9pt}

\begin{multicols}{2}

The method of molecular dynamics (MD) plays a prominent role in studying
molecular liquids. All existing techniques appropriate to simulate such
systems can be categorized in dependence on what type of parameters are
chosen to represent the rotational degrees of freedom and what kind of
numerical algorithm is applied to integrate the corresponding equations
of motion.

In the molecular approach, the phase trajectories are considered in view of
translational and rotational motions. The translational dynamics is defined
by motion of molecular centers of masses, whereas the orientational motion
can be determined in terms of Eulerian angles [1, 2], quaternions [3--8] or
principal-axis vectors [4]. The numerical integration within Eulerian angles
is very inefficient due to singularities of the equations of motion [3, 5].
If the quaternions or principal-axis vectors are involved, additional efforts
must be paid to conserve their unit norms or orthonormality.

The atomic approach [9] treats dynamics of the system in view of translational
motion of individual atoms which move under the potential-energy forces plus
forces of constraints introduced to hold inter-atomic distances constant. This
approach is believed to have good stability properties, because the usual
Verlet algorithm can be applied here. Nevertheless, the atomic approach is
sophisticated to implement for point molecules and when there are more than
two, three or four atoms in the cases of linear, planar and three-dimensional
molecules, respectively. Moreover, to reproduce the rigid molecular structure
it is necessary to solve complicated systems of nonlinear (in general, six per
molecule) equations at each time step of the integration [10].

It is a common practice to integrate orientational motion with the Gear
predictor-corrector algorithm of a high-order [11]. Such an algorithm,
being accurate at very small time steps, quickly becomes unstable with
increasing the step size [10]. Translational motion is usually integrated
with lower-order Verlet [12], velocity Verlet [13] and leapfrog [14]
integrators, owing their simplicity and exceptional numerical stability.
However, original versions of these integrators were constructed assuming
that acceleration is velocity-independent and, therefore, they can not be
applied directly to rotational dynamics. Analogous problems arise with
translational motion in the presence of magnetic fields.

In order to remedy that situation, Fincham [15] has derived a rotational-motion
version of the leapfrog algorithm in which systems of four nonlinear equations
per molecule for quaternion components are solved by iteration. Ahlrichs and
Brode have introduced a method [16] in which principal axes are considered
as pseudo-particles and constraint forces are introduced to maintain their
orthonormality. But the algorithm is within the Verlet framework and does
not contain angular velocities explicitly. The quaternion dynamics with
constraints was also formulated [17]. As a result, a new algorithm within
the velocity Verlet framework has been generated. Recently, the principal-axes
scheme has been adapted to this framework as well [18]. Nevertheless,
it was concluded that the best numerical stability can be achieved
in the atomic-constraint approach.

In this paper we propose a new leapfrog integrator of the rigid-body equations
of motion. The main idea consists in involving angular velocities, instead
of angular momenta, into the integration. This leads to significant
simplifications with respect to angular-momenta versions [15]. The algorithm
seems to be the most efficient and simple, exhibiting excellent stability
properties which are similar to those observed within the cumbersome
atomic-constraint technique.

Consider a classical system with $N$ rigid molecules composed of $M$ point
atoms. Translational motion of the system is described in the usual way,
applying Newton's law, whereas two first-order equations per molecule of
the rotational dynamics can be obtained as follows. According to Euler
equations [1], the rate of change in time of principal components,
$({\Omega}_{\rm X}^i,{\Omega}_{\rm Y}^i,{\Omega}_{\rm Z}^i)={\bf \Omega}_i$,
of angular velocity is
\begin{equation}
J_\alpha \frac{{\rm d} {\Omega}_\alpha^i}{{\rm d} t} = K_\alpha^i(t)
+ \left( J_\beta-J_\gamma \right) {\Omega}_\beta^i(t) {\Omega}_\gamma^i(t) .
\end{equation}
Here $(\alpha,\beta,\gamma)=({\rm X,Y,Z})$, $({\rm Y,Z,X})$ and $({\rm
Z,X,Y})$, $K_\alpha^i$ are principal components, ${\bf K}_i={\bf A}_i
{\bf k}_i^+$, of the torque ${\bf k}_i=\sum_{j;a,b}^{N;M} ({\bf r}_i^a-
{\bf r}_i) {\bf \times} {\bf f}_{ij}^{ab}$ exerted on molecule $i$ with
respect to its center of mass ${\bf r}_i$ due to the site-site interactions
${\bf f}_{ij}^{ab} \equiv {\bf f}({\bf r}_i^a-{\bf r}_j^b)$ with the other
molecules, $J_\alpha$ denote the principal moments of inertia, orientational
variables were collected into the square orthonormal matrices ${\bf A}_i$,
the nine elements of each of which ($i=1,\ldots,N$) present coordinates of
three principal axes (XYZ) of the molecule in the laboratory frame, the
position of atom $a$ within molecule $i$ in the same frame is ${\bf r}_i^a(t)
={\bf r}_i(t)+{\bf A}_i^+(t) {\bf \Delta}^a$, where ${\bf \Delta}^a=
(\Delta^a_1, \Delta^a_2, \Delta^a_3)^+$ is a vector-column of these
positions in the body frame and ${\bf A}^+$ the matrix transposed
to ${\bf A}$.

The second equation follows from definition of angular velocity,
\begin{equation}
\frac{{\rm d} {\bf A}_i}{{\rm d} t}=\left(
\begin{array}{ccc}
0 & {\Omega}_{\rm Z}^i & -{\Omega}_{\rm Y}^i \\
-{\Omega}_{\rm Z}^i & 0 & {\Omega}_{\rm X}^i \\
{\Omega}_{\rm Y}^i & -{\Omega}_{\rm X}^i & 0
\end{array}
\right)
{\bf A}_i \equiv {\bf W}({\bf \Omega}_i) {\bf A}_i ,
\end{equation}
where the property ${\bf A} {\bf A}^+={\bf I}$ of orthonormal matrices has
been used, ${\bf W}({\bf \Omega}_i)$ is a skewsymmetric matrix associated
with angular velocity, i.e., ${\bf W}^+({\bf \Omega}_i)=-{\bf W}({\bf
\Omega}_i)$ and ${\bf I}$ designates the unit matrix. In an alternative
representation the matrix ${\bf A}_i \equiv {\bf A} ({\bf q}_i)$ is a
function of the four-component quaternion ${\bf q}_i \equiv (\xi_i,\eta_i,
\zeta_i,\chi_i)^+$ [4, 5]. The time derivatives of quaternions can be cast
in the form
\begin{equation}
\frac{{\rm d} {\bf q}_i}{{\rm d}t}
= \displaystyle \frac12 \left(
\begin{array}{cccc}
0&{\Omega}_{\rm Z}^i&-{\Omega}_{\rm X}^i&-{\Omega}_{\rm Y}^i\\
-{\Omega}_{\rm Z}^i&0&-{\Omega}_{\rm Y}^i&{\Omega}_{\rm X}^i\\
{\Omega}_{\rm X}^i&{\Omega}_{\rm Y}^i&0&{\Omega}_{\rm Z}^i  \\
{\Omega}_{\rm Y}^i&-{\Omega}_{\rm X}^i&-{\Omega}_{\rm Z}^i&0
\end{array}
\right) {\bf q}_i \equiv {\bf Q}({\bf \Omega}_i) {\bf q}_i ,
\end{equation}
where ${\bf Q}({\bf \Omega}_i)$ is a skewsymmetric matrix again and the unit
quaternion norm $\xi_i^2+\eta_i^2+\zeta_i^2+\chi_i^2=1$, which follows from
the orthonormality of ${\bf A}_i$, has been used.

In the case of translational motion, it is easy to derive the leapfrog
algorithm [14]: ${\bf v}_i(t+\frac{h}{2})={\bf v}_i(t-\frac{h}{2})+h
{\bf a}_i(t)$, \, ${\bf r}_i(t+h)={\bf r}_i(t)+h{\bf v}_i(t+\frac{h}{2})$,
where $h$ denotes the time increment, ${\bf v}_i={\rm d} {\bf r}_i/{\rm d} t$
is the center-of-mass velocity, ${\bf a}_i(t)=\frac1m \sum_{j;a,b}^{N;M}
{\bf f}_{ij}^{ab}(t)$ the molecular acceleration and $m$ the mass of a
separate molecule. Recently, it has been shown that contrary to the
conventional point of view, the order of truncation errors for this
leapfrog is four rather than three for both coordinates and velocities
due to a fortunate cancellation of uncertainties [19].

The problems with deriving a leapfrog algorithm for rotational motion
are that angular accelerations (1) depend explicitly not only on spatial
coordinates via molecular toques but also on angular velocities. Moreover,
the time derivatives of orientational variables do not define angular
velocities directly (see Eqs. (2) and (3)). These difficulties can not be
handled with a simple leapfrog in which position and velocity are known
at different times. It is worth to underline that similar problems (even
much more difficult) arise in the angular-momentum approach [15], Verlet
and velocity Verlet frameworks [17, 18].

The basic idea of our approach lies in involving principal angular velocities
into the integration process. Then, acting in the spirit of leapfrog scheme
and using Euler equation (1), one obtains
\begin{eqnarray}
{{\Omega}_\alpha^i}^{\!(n+1)}(t+{\textstyle \frac{h}{2}})&=&
{\Omega}_\alpha^i(t-{\textstyle \frac{h}{2}})+
\frac{h}{J_\alpha} \Big[ K_\alpha^i(t) \nonumber \\ [-6pt]  \\ [-6pt]
&+& \left( J_\beta-J_\gamma \right) {{\Omega}_\beta^i}^{\!(n)}(t)
{{\Omega}_\gamma^i}^{\!(n)}(t) \Big] . \nonumber
\end{eqnarray}
While the molecular torques $K_\alpha^i(t)$ can easily be evaluated via
the coordinates ${\bf r}_i(t)$ and ${\bf A}_i(t)$ or ${\bf q}_i(t)$, a
propagation of the products of angular velocities in Eq. (4) to on-step
levels of time is necessary. The obvious choice for this is
\begin{eqnarray}
{{\Omega}_\beta^i}^{\!(n)}(t) {{\Omega}_\gamma^i}^{\!(n)}(t)&=&
\frac12 \Big[ {\Omega}_\beta^i(t-{\textstyle \frac{h}{2}})
{\Omega}_\gamma^i(t-{\textstyle \frac{h}{2}}) \nonumber \\ [-6pt]  \\ [-6pt]
&+& {{\Omega}_\beta^i}^{\!(n)}(t+{\textstyle \frac{h}{2}})
{{\Omega}_\gamma^i}^{\!(n)}(t+{\textstyle \frac{h}{2}}) \Big] . \nonumber
\end{eqnarray}
In view of (5), equation (4) constitutes a system of maximum three nonlinear
equations per molecule for the unknowns ${\Omega}_\alpha^i(t+{\textstyle
\frac{h}{2}})$. The system is simple and can be solved in a quite efficient
way by iteration, $n=0,1,\ldots,$ taking ${{\Omega}_\alpha^i}^{\!(0)}(t+
{\textstyle \frac{h}{2}})={\Omega}_\alpha^i(t-{\textstyle \frac{h}{2}})$
as an initial guess. We note that the order of truncation errors for
angular-velocity evaluation (4) reduces to three, because approximation
(5) is only second order accurate on $h$.

The evaluation of orientational variables can be realized by writing
\begin{equation}
{\bf S}_i(t+h)={\bf S}_i(t) + h {\bf H}_i
{\bf S}_i(t+{\textstyle \frac{h}{2}})
\end{equation}
for principal-axis vectors $({\bf S}_i \equiv {\bf A}_i, {\bf H}_i \equiv
{\bf W}_i)$ and quaternion $({\bf S}_i \equiv {\bf q}_i, {\bf H}_i \equiv
{\bf Q}_i)$ representations, where Eqs. (2) and (3) have been used. The
matrices ${\bf W}_i \equiv {\bf W} ({\bf \Omega}_i)$ and ${\bf Q}_i \equiv
{\bf Q}({\bf \Omega}_i)$ are calculated using already defined angular
velocities ${\bf \Omega}_i(t+{\textstyle \frac{h}{2}})$, whereas
orientational variables can be propagated to mid-step levels of time as
\begin{equation}
{\bf S}_i(t+{\textstyle \frac{h}{2}})={\textstyle \frac12}
\left[{\bf S}_i(t)+{\bf S}_i(t+h)\right] .
\end{equation}
Equation (6) together with (7) are, in fact, systems of linear equations
with respect to elements of ${\bf A}_i(t+h)$ and ${\bf q}_i(t+h)$, which,
therefore, can be solved analytically. The result is
\begin{equation}
{\bf S}_i(t+h)=
({\bf I}-{\textstyle {\textstyle \frac{h}{2}}} {\bf H}_i)^{-1}
({\bf I}+{\textstyle {\textstyle \frac{h}{2}}} {\bf H}_i) {\bf S}_i(t)
\equiv {\bf \Theta}_i(t,h) {\bf S}_i(t) .
\end{equation}
More explicit expressions for the set ${\bf \Theta}_i \equiv \{{\bf D}_i,
{\bf G}_i\}$ of evolution matrices are: ${\bf D}_i=[{\bf I}\,(1-\frac{h^2}{4}
\Omega_i^2)+h{\bf W}_i+\frac{h^2}{2}{\bf P}_i]/[1+\frac{h^2}{4} \Omega_i^2]$
and ${\bf G}_i=[{\bf I}\,(1-\frac{h^2}{16} \Omega_i^2)+h{\bf Q}_i]/[1+
\frac{h^2}{16} \Omega_i^2]$ in the cases of principal axes and quaternion
representations, respectively, where ${\bf P}_i$ is a symmetric matrix
with the elements ${\Omega}_\alpha^i(t+\frac{h}{2}) {\Omega}_\beta^i(t+
\frac{h}{2})$ and $\Omega_i^2 \equiv {\bf \Omega}_i^2(t+\frac{h}{2})$.
This completes the algorithm. It is interesting to remark that evaluation
(8) exhibits the same fourth-order local accuracy on $h$ as in the case of
translational coordinates, despite the second order of interpolation (7).
The reason for this results again from a cancellation of errors arising
in coordinates and velocities during two neighbor time steps.

It can be verified easily that the matrix $({\bf I}-\lambda {\bf H})^{-1}
({\bf I}+\lambda {\bf H})$ is orthonormal at arbitrary values of $\lambda$,
provided ${\bf H}^+=-{\bf H}$. Then, as follows from construction (8), the
evolution matrices ${\bf D}_i$ and ${\bf G}_i$ are orthonormal as well.
Therefore, if initially the orthonormality of ${\bf A}_i$ and unit norms
of ${\bf q}_i$ are satisfied, they will be fulfilled perfectly at arbitrary
times in future, despite the approximate character of produced trajectories.
This fact can be considered as the main advantage of the algorithm derived
that distinguishes it from all other singularity free algorithms, because no
additional efforts are needed to preserve the rigid structure of molecules.

We now test our approach on the basis of MD simulations on liquid water. The
simulations were performed in an NVE ensemble with $N=256$ molecules at a
density of $N/V$=1 g/cm$^3$ and at a temperature of 298 K using the TIP4P
potential ($M=4$) and reaction field geometry [20]. All runs were started
from an identical well equilibrated configuration. The numerical stability
was identified in terms of fluctuations of the total energy, ${\cal E}=
[\langle (E-\langle E \rangle)^2 \rangle ]^{1/2}/|\langle E \rangle|$. The
kinetic part of the energy was calculated at time $t$ putting ${\bf V}(t)=
\frac12 [{\bf V}(t-\frac{h}{2})+{\bf V}(t+\frac{h}{2})]+{\cal O}(h^2)$
for ${\bf V} \equiv \{{\bf v}_i,{\bf \Omega}_i\}$, where the main term
${\cal O}(h^2)$ of uncertainties is in the self-consistency with the
second order of global errors for our algorithm (one order lower than
minimal order of truncation errors for coordinates and velocities).

As the atomic-constraint algorithm [9, 10] is intensively exploited and its
performances are generally recognized, we have made comparative tests using
this method and our advanced leapfrog algorithm within quaternion and
principal-axes variables, as well as all known other approaches, namely,
the fifth-order Gear algorithm [11], implicit leapfrog of Fincham [15],
pseudo-particle formalism [16], quaternion- and matrix-constraint methods
[17,18]. Samples of ${\cal E}(t)$ as a function of the length of the
simulations at four fixed values of $h=$ 1, 2, 3 and 4 fs are shown in
Fig. 1. The usual value of step size for studying such a system is 2 fs [21].

Despite the Gear algorithm integrates the equations of motion very well at
$h=1$ fs, it has a very small region of stability and can not be used for
greater time steps (see Fig. 1 (b)). Small step sizes are impractical in
calculations because too much expensive computer time is required to
cover the sufficient phase space. At the same time, the quaternion- and
matrix-constraint methods as well as the pseudo-particle approach produce
much more stable trajectories and exhibit similar equivalence in the energy
conservation. Worse results are observed for the Fincham's leapfrog method.
Finally, the best numerical stability is achieved in the atomic-constraint
technique and our leapfrog scheme within both quaternion and principal axes
representations, which conserve the energy approximately with the same
accuracy (the results for principal-axis variables and pseudo-particle
formalism are not included in the figure to simplify graph presentation).
Quite a few iterations (the mean number of iterations varied from 3 to 5
at $h = 1 \div 4$ fs) was sufficient to find solutions to the system of
nonlinear equations (4) with a precision of $10^{-12}$. This contributes
a negligible small computation time additionally into the total time.

No shift of the total energy was observed for the atomic-constraint and our
leapfrog techniques at $h \le 4$ fs over a length of 10 000 steps. To
reproduce features of an NVE ensemble quantitatively, it is necessary for
the ratio $\Gamma={\cal E}/\Upsilon$ of the total energy fluctuations to the
fluctuations $\Upsilon$ of the potential energy to be no more than a few per
cent. We have obtained the following levels of ${\cal E}$ at the end of the
runs in our leapfrog approach: 0.0016, 0.0065, 0.015 and 0.029 \%,
corresponding to $\Gamma \approx$ 0.29, 1.2, 2.7 and 5.2 \% at $h$= 1, 2, 3
and 4 fs, respectively (for the system under consideration $\Upsilon \approx
0.56 \%$). Therefore, the greatest time step considered (4 fs) is still
suitable for precise calculations. The ratio $\Gamma$ can be fitted with a
great accuracy to the function $C h^2$ with a coefficient of $C \approx
0.29$ \% fs$^{-2}$. This is completely in line with our theoretical
prediction about a characteristic square growth of global errors and,
as a consequence, ${\cal E}(t)$ at $t \gg h$. The square growth was
observed in all other approaches, excepting the Gear algorithm. However,
only the advanced leapfrog algorithm provides a minimum of $C$ and total
energy fluctuations.

The algorithm presented might become popular because of its great stability,
simplicity to implement for arbitrary rigid bodies and its intrinsic
conservation of rigid structures. These features should be considered
as significant benefits of the algorithm with respect to all the rest
approaches. It can easily be substituted into existing MD programs on
rigid polyatomics. Moreover, since velocities appear explicitly, the
algorithm can be extended to a thermostat version and to integration
in the presence of magnetic fields. These problems will be discussed
in a separate publication.

The author thanks the President of Ukraine for financial support.

\vspace{24pt}

\begin{center}
{\large Figure caption}
\end{center}

Fig. 1. The total energy fluctuations as functions of the length of the
simulations on liquid water, performed in various techniques at four
fixed time steps: (a) 1 fs, (b) 2 fs, (c) 3 fs and (d) 4 fs.

\end{multicols}

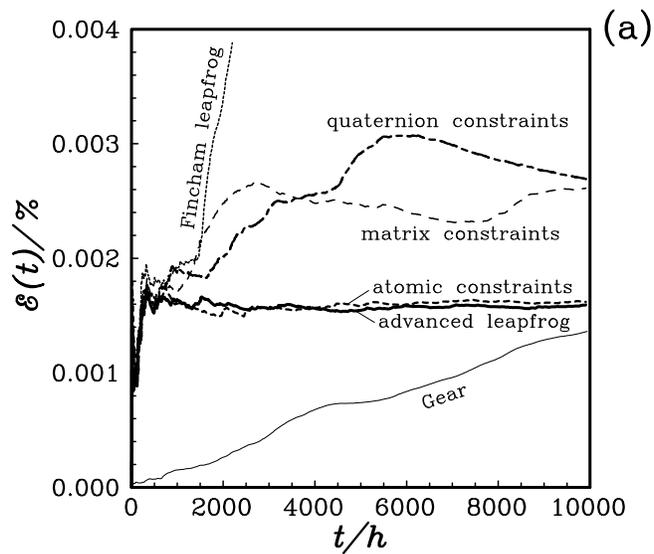
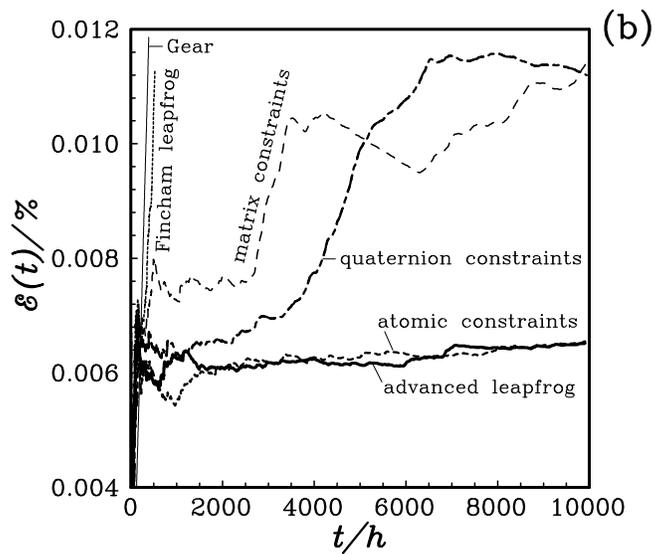
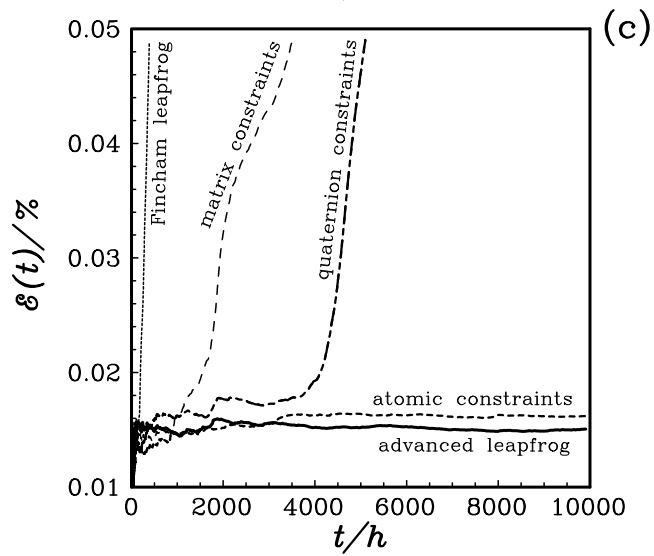
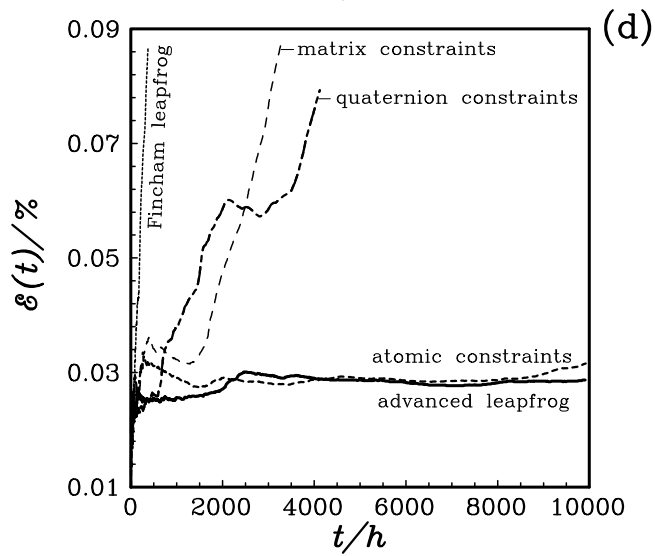
\end{document}